\documentclass[11pt]{cernrep}
\usepackage{graphicx}
\usepackage{colortbl}

\newcommand{\Pt}{{P_t}}

\newcommand{\dphi}{\Delta\phi}
\newcommand{\phizj}{\phi_{(Z,jet)}}

\newcommand{\ptzj}{$~\Pt^Z$ and $\Pt^{jet}~$}

\newcommand{\zpj}{~``$Z^0+jet$''~}
\newcommand{\gpj}{~``$\gamma^{dir}+jet$''~}

\newcommand{\pth}{\hat{p}_{\perp}^{\;min}}

\newcommand{\Ptz}{\Pt^{Z}}

\newcommand{\hmm}{\hspace*{-1.3mm}}

\newcommand{\lt}{\!<\!}

\suppressfloats[!]

\begin{document}

\begin{center}
{\bfseries \large On the application of ``$Z^0+jet$'' events for  determining the gluon distribution 
in a proton at the LHC.}

\vskip 5mm

D.V.~Bandurin$^1$, ~~N.B.~Skachkov$^2$

\vskip 5mm

{\small 
{\it ~Joint Institute for Nuclear Research, Dubna, Russia}\\
{\it
E-mail: (1) dmv@cv.jinr.ru, (2) skachkov@cv.jinr.ru
}}
\end{center}

\vskip 5mm
\begin{center}
\begin{minipage}{150mm}
\centerline{\bf Abstract}
\noindent
It is shown that the samples of \zpj events, collected at the LHC with the integrated luminosity 
$L_{int}=20~fb^{-1}$,
may have enough statistics for determining the gluon distribution inside a proton in the region 
of $2\cdot 10^{-4}\leq x \leq 1.0$ at $Q^2$ values in the interval of
$0.9\cdot10^3\leq Q^2\leq 4\cdot 10^4 ~(GeV/c)^2$. A possibility of the background events
suppression by use of the \zpj events selection criteria is also demonstrated.
%
\end{minipage}
\end{center}

\thispagestyle{empty}

\vskip 9mm

\setcounter{page}{1}
\section{Introduction.} 

Many important predictions for the production processes of new particles 
at the LHC  require a good knowledge of the gluon distribution function in a proton $f^p_g(x,Q^2)$. 
Thus, determining the proton gluon density 
directly in the LHC experiments, especially in the region of small  $x$ and high $Q^2$,
would be very useful. 

One of promising channels that can be used for measuring $f^p_g(x,Q^2)$ is 
the direct photon production process in association with one jet
$pp\rightarrow \gamma^{dir}\, +\, 1\,jet\, + \,X$.
It was studied in detail in \cite{BKS_GLU,BALD00,GLU_BKGD}
\footnote{see also \cite{MD1}}. 

Here for the same aim 
we consider the \zpj production process (see also \cite{MD1,Wom}), 
analogous to the \gpj process above: \\[-12pt]
\begin{equation}
pp\rightarrow Z^0\, +\, 1\,jet\, + \,X.
\label{eq:zpj}
\end{equation}
The process (\ref{eq:zpj}) is caused at the parton level  by two subprocesses: 
Compton-like scattering\\[-7mm]
\begin{eqnarray}
\hspace*{6.34cm} qg\to q+Z^0 \hspace*{70mm} (2a)
\nonumber
\end{eqnarray}
\vspace{-3mm}
and the annihilation process\\[-5mm]
\begin{eqnarray}
\hspace*{6.32cm} q\overline{q}\to g+Z^0.  \hspace*{70mm} (2b)
\nonumber
\end{eqnarray}
Here we suppose that $Z^0$ boson decays in the following via leptonic channels $Z^0\to \mu^+\mu^-,e^+e^-$,
the signals from which can be well reconstructed by using electromagnetic calorimeter, tracker and muon system
\cite{TP,TR,MS,CMS_zpj}
\footnote{The estimations done here are based on the geometry of the CMS detector \cite{TP}.}.

\setcounter{equation}{2}

In the case of $pp\to Z^0/\gamma^{dir}+1~ jet+X$  in the region of $\Pt^{Z/\gamma}\geq 30~GeV/c$ 
(where $k_T$ smearing
effects are not important \cite{Hu2}) the cross section of \zpj production is
expressed {\it directly}
\footnote{In contrast to, for instance, the cross section of the inclusive photon production
process, also used for the extraction of data on  $f^p_g(x,Q^2)$, that is expressed as integral 
over the proton momentum fractions $x_a$ multiplied by $f^p_a(x,Q^2)$.}
in terms of parton distribution functions $f_a^p(x_a,Q^2)$ and 
the cross sections of the elementary scattering subprocesses (e.g. see \cite{Owe}): \\[-15pt]
\begin{eqnarray}
\frac{d\sigma}{d\eta_1d\eta_2d\Pt^2} = \sum\limits_{a,b}\,x_a\,f_a^p(x_a,Q^2)\,
x_b\,f_b^p(x_b,Q^2)\frac{d\sigma}{d\hat{t}}(a\,b\rightarrow 1\,2)
\label{eq:cross_gl}
\end{eqnarray}
\noindent
where the incident parton momentum fractions $x_{a,b}$ can be found from the $Z^0$ and jet
parameters via \\[-22pt]
\begin{eqnarray}
x_{a,b} \,=\,\Pt/\sqrt{s}\cdot \,(exp(\pm \eta_{1})\,+\,exp(\pm \eta_{2})).
\label{eq:x_gl}
\end{eqnarray}
We also used the following designations above:
$\eta_1=\eta^Z$, $\eta_2=\eta^{jet}$; ~$\Pt=\Ptz$;~ $a,b = q, \bar{q},g$; 
$1,2 = q,\bar{q},g,Z^0$.
Formula (\ref{eq:cross_gl}) and the knowledge of the results of independent measurements of
$q, \,\bar{q}$ distributions \cite{MD1} allow the gluon  distribution $f^p_g(x,Q^2)$
to be determined in different $x$ and $Q^2$ intervals after a suppression of the background events contribution.

\section{Definition of selection rules.}

\noindent
1. We shall select the events with $Z^0$ boson 
\footnote{Here and below in the paper speaking about $Z^0$ boson we imply a signal
reconstructed from the lepton pair with leptons selected by the criteria $2-4$ of this section.}
and  one jet with\\[-5pt]
\begin{equation}
\Pt^{Z} \geq 30~ GeV/c~ \quad {\rm and} \quad \Pt^{jet}\geq 25 \;GeV/c.
\label{eq:sc1}
\end{equation}

The jet is defined according to the PYTHIA \cite{PYT}
jetfinding algorithm LUCELL having
the cone radius  counted from the jet initiator cell ($ic$)  
$R_{ic}=((\Delta\eta)^2 + (\Delta\phi)^2)^{1/2}=0.7$.
A jet pseudorapidity $|\eta^{jet}|$ is limited by $5.0$ according to the CMS detector geometry. 

\noindent
2. To guarantee a clear track identification of a lepton from the decays of $Z^0\to \mu^+\mu^-,e^+e^-$  
in the tracker and muon systems and most precise 
determination of its parameters we put the following restrictions on leptons
\footnote{Most of the $e,\mu$ selection cuts are taken from \cite{TR,MS,CMJ}.}:
~\\[-5mm]

(a) on the transverse momentum value $\Pt^l$ of any considered lepton:\\[-7pt]
\begin{equation}
\Pt^l \geq 10 ~GeV/c;
\label{eq:sc2a}
\end{equation}

\vskip-2.0mm
(b) on the $\Pt$ value of the most energetic lepton in a pair:\\[-5pt]
\begin{equation}
\Pt^l_{max} \geq \Pt^l_{CUT}
\label{eq:sc3b}
\end{equation}
This cut depends on the energy scale \cite{CMS_zpj}.
So, we have taken $\Pt^l_{CUT}= 20~GeV/c$ for events with $\Ptz\geq 40~GeV/c$ and
$\Pt^l_{CUT}= 50~GeV/c$ for events with $\Ptz\geq 100~GeV/c$.\\[-5mm]

(c) on the value of the ratio of $\Pt^{isol}$, i.e. 
the scalar sum of $\Pt$ of all particles surrounding a lepton, to $\Pt^l$ ($\Pt^{isol}/\Pt^l$) 
in the cone of radius $R=0.3$ and
on the value of maximal $\Pt$ of a charged particle surrounding a lepton in this cone:\\[-10pt]
\begin{equation}
\Pt^{isol}/\Pt^l \leq 0.10, \quad \Pt^{ch} \leq 2 ~GeV/c.
\label{eq:sc3c}
\end{equation}
The isolated high-$\Pt$ tracks (what takes place in case of the leptonic $Z^0$ decays)
should be reconstructed with a higher efficiency 
and with generation of a lower number of fake and ghost tracks \cite{TR,MS}.

\noindent
3. A lepton is selected in the acceptance region \cite{TR,MS} :\\[-10pt]
\begin{equation}
 |\eta^l|<2.4.
\end{equation}
\noindent
4. To select lepton pairs only from  $Z^0$ decay we limit the value of the lepton pair invariant mass
$M_{inv}^{ll}$ by
\footnote{A narrower mass window can be used with the statistics growth.}:
\begin{equation}
|M^Z - M_{inv}^{ll}| \leq 5 ~GeV/c^2.
\label{eq:sc3}
\end{equation}
with $M^Z$ taken to be $91.2~GeV/c^2$.

\noindent  
5. We select the events with the vector $\vec{\Pt}^{jet}$ being ``back-to-back" to
the vector $\vec{\Pt}^{Z}$ (in the plane transverse to the beam line)
within the azimuthal angle interval $\dphi$ defined by the equation:\\[-5pt]
\begin{equation}
\phizj=180^\circ \pm \Delta\phi 
\label{eq:sc4}
\end{equation}
where $\phizj$ is the angle between the $\vec{\Pt}^Z$ and $\vec{\Pt}^{jet}$ vectors: 
$\vec{\Pt}^{Z}\vec{\Pt}^{jet}=\Pt^{Z}\Pt^{jet} cos(\phizj)$, ~~~
with ~$\Pt^{Z}=|\vec{\Pt}^{Z}|,~~\Pt^{jet}=|\vec{\Pt}^{jet}|$. 
Here we limit $\Delta\phi$ values by $15^\circ$.

\noindent
6. The initial and final state radiations manifest themselves most clearly
as some final state mini-jets or clusters activity \cite{CMS_zpj,BKS,GPJ_D0}. 
To suppress it, we impose a new cut condition that 
was not formulated in an evident form in previous experiments: we select the \zpj events
that do not have any other jet-like or cluster high $\Pt$ activity  by taking values of
$\Pt^{clust}$ (with the cluster cone of $R_{clust}=0.7$), being smaller than some threshold
$\Pt^{clust}_{CUT}$ value, i.e. we select the events with\\[-7pt]
\begin{equation}
\Pt^{clust} \leq \Pt^{clust}_{CUT}.
\label{eq:sc5}
\end{equation}

\noindent
7. We limit the value of the modulus of the vector sum of $\vec{\Pt}$ of all
particles that do not belong to the \zpj system but 
fit into the region $|\eta|\lt5$ covered by
the calorimeter system, i.e., we limit the signal in the cells ``beyond the jet and $Z^0$'' regions
by the following cut:\\[-7pt]
\begin{equation}
\left|\sum_{i\not\in jet,Z^0}\vec{\Pt}^i\right| \equiv \Pt^{out} \leq \Pt^{out}_{CUT},
~~|\eta^i|\lt5.
\label{eq:sc6}
\end{equation}

\noindent
The importance of $\Pt^{out}_{CUT}$ and $\Pt^{clust}_{CUT}$
for selection of events with a good balance of \ptzj was already shown in
\cite{CMS_zpj,BKS,GPJ_D0}. In this paper they are fixed as $\Pt^{out}_{CUT}=10~GeV/c$
and  $\Pt^{out}_{CUT}=10~GeV/c$.



As we show below the presented selection criteria  
guarantee practically a complete suppression of the background events. 

\section{The study of background suppression.}                        

In principle, there is a probability, that some combination of $\mu^+\mu^-$ or $e^+e^-$ pairs in
the events, based on the QCD subprocesses with much larger cross sections 
(by about 5 orders of magnitude) than ones of  
the signal subprocesses (2a) and (2b),
can be registered as the $Z^0$ signal. 

Firstly, to study a rejection possibility of such type of events by about 40 million events
with a mixture of all QCD and SM subprocesses with large
cross sections existing in PYTHIA 
\footnote{Namely, having ISUB=11--20, 28--31, 53, 68 in PYTHIA \cite{PYT}.} 
including also the signal subprocesses 
\footnote{ISUB=15 and 30 in PYTHIA \cite{PYT}.} 
were generated with the only $Z^0$ decay mode allowed:  $Z^0\to \mu^+\mu^-$. 
Three generations were performed with different minimal $\Pt$ of the hard $2\to 2$ subprocess
\footnote{i.e. CKIN(3) parameter in PYTHIA \cite{PYT}.}
$\pth$ values: $\pth$= 40, 70 and 100 $GeV/c$. 
The cross sections of different subprocesses 
serve in simulation as weight factors and, thus, determine
the final statistics of the corresponding physical events.
The generated events were analyzed by use of the cuts given in Table \ref{tab:sb_mc0}
(see also Section 2)
\footnote{Notice, that the cuts used in Table \ref{tab:sb_mc0} are weak enough.
For instance, they do not limit (directly) $\Ptz$, $\Pt$ of the most energetic lepton in the pair
(as well as they do not include $\Pt^{clust}_{CUT}$ and $\Pt^{out}_{CUT}$).}.
\\[-5mm]
\begin{table}[h]
\small
\caption{List of the applied cuts used in Tables \ref{tab:sb_mc1}, \ref{tab:sb_mc2}.}
\begin{tabular}{lc} \hline
\label{tab:sb_mc0}
~~ \hspace*{148mm} ~~\\[-3mm]
\hspace*{14mm} {\bf 0}. Total number of $l^+ l^-$-- pairs (No selection); ~~~~~~~~~~~~~~~~~~~~~~~~~~~~~~~~~~~\\[1pt]
\hspace*{15mm}{\bf 1}. $\Pt^{l}>10~~GeV/c$, $|\eta^{l}|<2.4$; \\[1pt]   
\hspace*{15mm}{\bf 2}. $|M_{\bf Z}-M_{inv}^{ll}|<20~~GeV/c^2$; \\[1pt]
\hspace*{15mm}{\bf 3}. 1 jet events selected; \\[1pt]
\hspace*{15mm}{\bf 4}. $\Pt^{isol}/\Pt^l \leq 0.10,~\Pt^{ch}<2 ~~GeV/c $;\\[1pt]
\hspace*{15mm}{\bf 5}. $|M_{\bf Z}-M_{inv}^{ll}|<5~~GeV/c^2$; \\[1pt]
\hspace*{15mm}{\bf 6}. $\dphi<15^\circ$. \\\hline 
\end{tabular}
\end{table}

To trace the effect of their application let us consider first the case of one
(intermediate) energy, i.e. the generation with $\pth$=70 $GeV/c$.
Each line of Table \ref{tab:sb_mc1} corresponds to the respective cut of Table \ref{tab:sb_mc0}.
The numbers in columns ``Signal'' and ``Bkgd'' show the number of muon pairs in the signal and (combinatorial)
background events remained after a cut. Column ``$Eff_{S(B)}$'' demonstrates the efficiency 
of a cut.
The efficiencies $Eff_{S(B)}$ (with their errors) are defined as a ratio
of the number of signal (background) events that passed under a cut
(1--6) to the number of the preselected events after the first cut of Table \ref{tab:sb_mc0}
\footnote{The number of events after the first cuts is taken as $100\%$.}.
\begin{table}[h]
\small
\begin{center}
\vskip-1mm
\caption{A demonstration of cut-by-cut efficiencies and $S/B$ ratios for generation with
$\pth$=70 $GeV/c$}
\hspace*{0mm} \footnotesize{($Z^0\to \mu^+\mu^-$).}
\small
\vskip0.1cm
\label{tab:sb_mc1}
\begin{tabular}{|c||c|c|c|c|c|}                  \hline 
Selection & Signal & Bkgd & $Eff_S(\%)$ &$Eff_B(\%)$ & $S/B$ \\\hline 
 0  &  401 & 850821 &           &             & $5\cdot10^{-4}$ \\\hline \hline  
 1  &  245 &  15842 &100.00$\pm$0.00 &100.00$\pm$0.000 & 0.02 \\\hline 
 2  &  226 &    467 & 92.24$\pm$8.51 &  2.948$\pm$0.138 & 0.5 \\\hline 
 3  &   99 &     12 & 40.41$\pm$4.81 &  0.076$\pm$0.022 & 8.3 \\\hline 
 4  &   81 &     10 & 33.00$\pm$4.24 &  0.063$\pm$0.020 & 8.1 \\\hline 
 5  &   72 &      4 & 29.39$\pm$3.94 &  0.025$\pm$0.013 &18.0 \\\hline 
 6  &   62 &      0 & 25.31$\pm$3.60 &  0.000$\pm$0.000 & --  \\\hline         
\end{tabular}
\end{center}
\end{table}

We see from Table \ref{tab:sb_mc1} that initial ratio of $\mu^+\mu^-$ pairs in
signal and background events is very small ($5\cdot10^{-4}$)
\footnote{That is mainly due to the huge difference in the cross sections of \zpj events (from subprocesses
(2a), (2b)) and the QCD events.}.
%
A weak restriction of the muon transverse momentum and pseudorapidity
in the first selection increase $S/B$ by about 2 order 
(as $5\cdot10^{-4}\to2\cdot10^{-2}$).
The invariant mass criterion and one-jet events selection make $S/B=18.0$ and
the last criterion on the azimuthal angle between $Z^0$ and jet 
($\dphi<15^\circ$) suppresses the background events completely.

The information on other intervals (i.e. on the event generations with $\pth=40$ and $\pth=100~GeV/c$)
is presented in Table \ref{tab:sb_mc2}.
Line ``Preselection (1)'' corresponds to the first cuts in Table \ref{tab:sb_mc0} 
($\Pt^{\mu}>10~~GeV/c$, $|\eta^{\mu}|<2.4$) while line ``Main ($1-5$)'' corresponds to the result of
application of criteria from 1 to 5 of Table \ref{tab:sb_mc0}. 
{\it After application of all six criteria of Table \ref{tab:sb_mc0} 
we have observed no background events in all of the $\Ptz$ intervals} 
with the signal events selection efficiency of $25-33\%$.

Analogous simulations in PYTHIA were done to estimate a background
to the \zpj events with the subsequent $Z^0$ decay via $e^+e^-$ channel. By about 20 million events
were generated at $\pth$= 40, 70 and 100 $GeV/c$
with a mixture of all QCD and SM subprocesses. The results are given in Table \ref{tab:sb_ec2}.
As in the case of $Z^0\to \mu^+\mu^-$, 
no background events were found after application of all criteria of Table \ref{tab:sb_mc0}.
\begin{table}[h]
\small
\begin{center}
\vskip-1mm
\caption{Values of efficiencies and $S/B$ ratios for generations with
$\pth$=40, 70 and 100 $GeV/c$}
\hspace*{5mm} \footnotesize{($Z^0\to \mu^+\mu^-$).}
\small
 \vskip0.1cm
\label{tab:sb_mc2}
\begin{tabular}{|c||c|c|c|c|c|c|}                  \hline 
$\pth$& Cuts& Signal& Bkgd &$Eff_S(\%)$&$Eff_B(\%)$&$S/B$\\\hline \hline
40 &Preselection (1)  & 89&  1090&100.00$\pm$0.00 &100.00$\pm$0.00 &0.08\\\cline{2-7}
$(GeV/c)$ & Main ($1-5$) & 30&     0& 33.71$\pm$7.12 & 0.000$\pm$0.000  & -- \\\hline \hline
70 &Preselection (1)  &245& 15842&100.00$\pm$0.00 &100.00$\pm$0.00  &0.02\\\cline{2-7}
$(GeV/c)$ & Main ($1-5$) & 72&     4& 29.39$\pm$3.94 & 0.025$\pm$0.013 &18.0\\\hline \hline
100 &Preselection (1) &497& 37118&100.00$\pm$0.00 &100.00$\pm$0.00  &0.01\\\cline{2-7}
$(GeV/c)$ & Main ($1-5$) &127&     4& 25.55$\pm$2.54 & 0.011$\pm$0.005 &31.8\\\hline 
\end{tabular}
\end{center}
\vskip-9mm
\end{table}

\begin{table}[h]
\small
\begin{center}
\caption{Values of efficiencies and $S/B$ ratios for generations with
$\pth$=40, 70 and 100 $GeV/c$}
\hspace*{5mm} \footnotesize{($Z^0\to e^+e^-$).}
\small
 \vskip0.2cm
\label{tab:sb_ec2}
\begin{tabular}{|c||c|c|c|c|c|c|}                  \hline 
$\pth$& Selections& Signal& Bkgd &$Eff_S(\%)$&$Eff_B(\%)$&$S/B$\\\hline \hline
40 &Preselected (1)  & 48&  1404&100.00$\pm$0.00 &100.00$\pm$0.00 &0.03\\\cline{2-7}
$(GeV/c)$ & Main ($1-5$) & 20&     3& 41.67$\pm$11.09 & 0.214$\pm$0.123  & 6.7 \\\hline \hline
70 &Preselected (1)  &95& 5396&100.00$\pm$0.00 &100.00$\pm$0.00  &0.02\\\cline{2-7}
$(GeV/c)$ & Main ($1-5$) & 35&     2& 36.32$\pm$7.32 & 0.037$\pm$0.026 &17.5\\\hline \hline
100 &Preselected (1) &191& 18158&100.00$\pm$0.00 &100.00$\pm$0.00  &0.01\\\cline{2-7}
$(GeV/c)$ & Main ($1-5$) &61&     2& 31.68$\pm$4.67 & 0.008$\pm$0.007 &30.5 \\\hline 
\end{tabular}
\end{center}
\end{table}

The practical absence of a background to the \zpj events allow
to use them for an extraction of the gluon distribution in a proton $f^p_g(x,Q^2)$.

%
\section{Estimation of rates for gluon distribution determination.}
%

In Table~\ref{tab:B30} we present the distribution  of the number of the events,
 based on the subprocesses $qg\to Z^0+q$ and $q\bar{q}\to g+Z^0$ 
(with the decays $Z^0\to\mu^+\mu^-, e^+e^-$), at integrated luminosity $L_{int}=20 ~fb^{-1}$
in different $x$ (defined by (\ref{eq:x_gl})) and $Q^2 (\equiv(\Ptz)^2)$ intervals.
These events satisfy the cuts (5)--(13) of Section 2. We see that 
at  $L_{int}=20 ~fb^{-1}$ one can collect about half a million of
\zpj events in the interval of $30\leq\Ptz \leq 200 ~GeV/c$.
\begin{table}[htbp]
\small
\begin{center}
\caption{Numbers of \zpj events (with $Z^0\to\mu^+\mu^-, e^+e^-)$ in 
$Q^2$ and $x$ intervals for $L_{int}=20 ~fb^{-1}$.}
\label{tab:B30}
\vskip0.1cm
\begin{tabular}{|lc|r|r|r|r|r|c|}                  \hline
 & $Q^2$ &\multicolumn{4}{c|}{ \hspace{-0.9cm} $x$ values of a parton} &All $x$ 
&$\Pt^{Z}$   \\\cline{3-7}
 & $(GeV/c)^2$ & $10^{-4}$--$10^{-3}$ & $10^{-3}$--$10^{-2}$ &$10^{-2}$--
$10^{-1}$ & $10^{-1}$--$10^{0}$ & $10^{-4}$--$10^{0}$&$(GeV/c)$     \\\hline
&\hmm\hmm  900-1600\hmm  & 36818  & 91689  & 94905  &  4957  &228369 & 30--40\\\hline 
&\hmm\hmm 1600-2500\hmm  & 14833  & 56722  & 57403  &  3708  &132667 & 40--50\\\hline
&\hmm\hmm 2500-3600\hmm  &  4957  & 33148  & 38029  &  3065  & 79199 & 50--60\\\hline
&\hmm\hmm 3600-5000\hmm  &  2195  & 20812  & 25882  &  3065  & 51954 & 60--71 \\\hline
&\hmm\hmm 5000-6400\hmm  &   454  & 11693  & 13887  &  2043  & 28077& 71--80\\\hline
&\hmm\hmm 6400-8100\hmm  &   341  &  8476  & 10860  &  1249  & 20925& 80--90\\\hline
&\hmm\hmm 8100-10000\hmm &    38  &  5979  &  8098  &  1438  & 15552& 90--100\\\hline
&\hmm\hmm 10000-14400\hmm&    38  &  5638  &  9157  &  1816  & 16650& 100--120\\\hline
&\hmm\hmm 14400-20000\hmm&     0  &  2800  &  5562  &   908  &  9271& 120--141\\\hline
&\hmm\hmm 20000-40000\hmm&     0  &  1816  &  4389  &  1438  &  7644& 141--200 \\\hline
\multicolumn{6}{c|}{}&{\bf 590~308}\\\cline{7-7}
\end{tabular}
\end{center}
\end{table}

\begin{figure}[htbp]
\vskip-7mm
   \hspace{10mm} \includegraphics[width=.90\linewidth,height=79mm,angle=0]{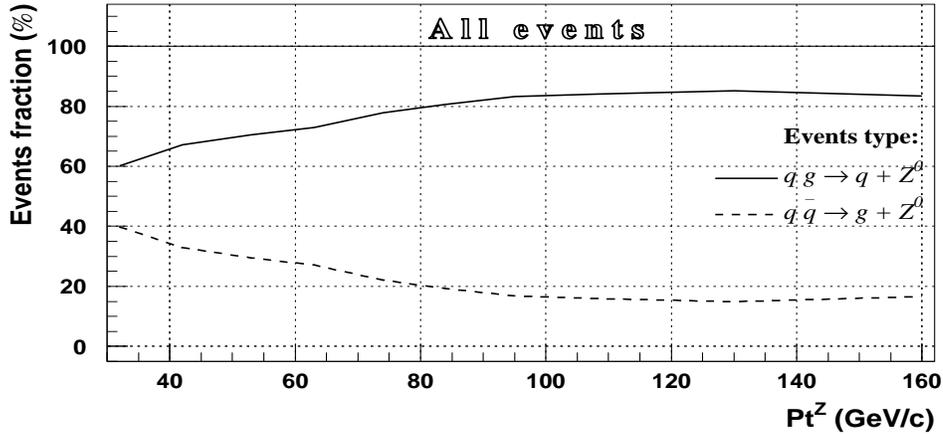}
\vskip-9mm
\caption{The contributions of the events originated from the subprocesses (2a) and (2b) 
as a function of $\Ptz$. Full line corresponds to the ``$qg\to q+Z^0$'' events, 
dashed line -- to the ``$q\bar{q}\to g+Z^0$'' events.}
\label{fig:procs}
\end{figure}

The contributions (in $\%$) of the events originated from the subprocesses (2a) and (2b) 
\footnote{and passed  selection cuts (5)--(13)}
as functions of $\Ptz$ are presented in Fig.~\ref{fig:procs}. 
From this figure one can see that the fraction of the ``gluonic'' events originated from the Compton
scattering (2a) noticeably dominates over all considered $\Ptz$ interval and
varies from about $60\%$ at $\Ptz\approx 30~GeV/c$ to about $85\%$ at $\Ptz\geq 100~GeV/c$.

\begin{figure}[h]
   \vskip-2mm
   \hspace{37mm} \includegraphics[width=.53\linewidth,height=7.9cm,angle=0]{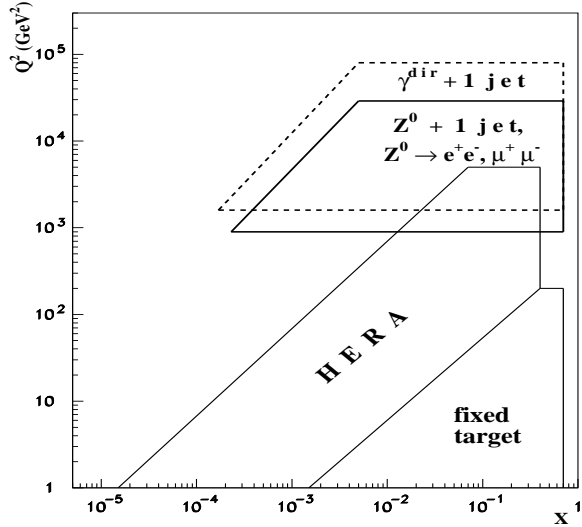}
\vspace*{-5.0mm}
\caption{LHC  $(x,Q^2)$ kinematic region for the process
$pp\to Z^0 + jet+X$ ~~(with $Z^0\to \mu^+\mu^-, e^+e^-)$.}
\label{fig:zpj_xQ2}
  \vskip2mm
\end{figure}

The $x-Q^2$ kinematic area in which one can study the gluon distribution $f^p_g(x,Q^2)$
by selecting \zpj events (with the  leptonic decay modes of $Z^0$)
is also shown in Fig.~\ref{fig:zpj_xQ2}.
From this figure (and Tables~\ref{tab:B30}) 
it is seen that during first two years of LHC running 
at low luminosity ($L=10^{33}\,cm^{-2} s^{-1}$) it would be possible to extract an information
for  determination of $f^p_g(x,Q^2)$ in the region of 
$0.9\cdot10^3\leq Q^2\leq 4\cdot 10^4 ~(GeV/c)^2$ with as small $x$ values as accessible at HERA
but at higher $Q^2$ values (by 1--2 orders of magnitude).
It is also worth emphasizing that the  sample of the \zpj events selected for this aim can be used
to perform a cross-check of $f^p_g(x,Q^2)$ determination by using  \gpj events 
\cite{BKS_GLU,BALD00,GLU_BKGD}.
It is especially important in the region of lower $Q^2$ where
we have quite a sufficient statistics of \zpj events, on the one hand, and a higher
background contribution to the \gpj events, on the other hand.
The area that can be covered with \gpj events is also shown in Fig.~\ref{fig:zpj_xQ2} by dashed lines.

~\\[2mm]

\section{Summary.}

It is shown that the  samples of \zpj events with a clean topology, 
most suitable for the absolute jet energy  scale setting 
\footnote{As was shown in \cite{CMS_zpj}, the chosen cut conditions noticeably suppress 
initial and final state radiations, i.e. the contributions from the events caused by next-to-leading order diagrams}
\cite{CMS_zpj} and with the suppressed combinatorial background contribution from the QCD events, 
can provide an useful information
for the gluon density determination inside a proton.
The corresponding measurements can be done in a new kinematic region, not covered in any previous
experiments, of $2\cdot 10^{-4}\leq x \leq 1.0$ with $0.9\cdot10^3\leq Q^2\leq 4\cdot 10^4 ~(GeV/c)^2$.
The study of gluon distribution $f^p_g(x,Q^2)$ obtained from the analysis of \zpj events  can be used
as the independent cross-check of the $f^p_g(x,Q^2)$ determination from the \gpj events 
\cite{BKS_GLU,BALD00,GLU_BKGD} as well as from the analytical solutions of the DGLAP equations describing
the $Q^2$ evolution of parton distributions at small $x$ \cite{Kot}.

~\\[5mm]

\noindent
{\bf Acknowledgments}
\newline
We thank P.~Aurenche, D.~Denegri, M.Dittmar, M.~Fontannaz, J.Ph.~Guillet, M.L.~Mangano, 
E.~Pilon and S.~Tapprogge for helpful discussions.


\begin{thebibliography}{99}
\bibitem{BKS_GLU}
 D.V.~Bandurin, V.F.~Konoplyanikov, N.B.~Skachkov,
``\gpj events rate estimation for gluon distribution determination at LHC'',
Part.Nucl.Lett.{\bf 103}:34-43,2000, hep-ex/0011015.
\bibitem{BALD00}
 D.V.~Bandurin, V.F.~Konoplyanikov, N.B.~Skachkov,
``Events rate estimation for gluon distribution determination at LHC'', hep-ex/0207028.
Proc. of the XV ISHEP
``Relativistic Nuclear Physics and Quantum Chromodynamics'',
Dubna 2000. Eds.A.M.~Baldin, V.V.~Burov, A.I.~Malakhov. Dubna, 2001, v.I, pp.375-383.
\bibitem{GLU_BKGD}
D.V.~Bandurin, N.B.~Skachkov,
``On the possibility of measuring the gluon distribution in proton with
``$\gamma+jet$'' events at LHC'', hep-ex/0210004 (To appear as CMS Note).
\bibitem{MD1} 
M.~Dittmar, F.~Pauss, D.~Zurcher, Phys.Rev. {\bf D56} (1997)7284.
\bibitem{Wom}
J.~Womersley, A talk at CMS Week meeting, Aachen, 1997.
\bibitem{Hu2} 
 J.~Huston ATLAS Note ATL-Phys-99-008, CERN,1999.
\bibitem{TP}
CMS, Technical proposal, CERN/LHCC 94-38.
\bibitem{TR}
The CMS Tracker Project, CERN/LHCC 98--6, CMS TDR 5, CERN, 1999.      
\bibitem{MS}
The CMS Muon Project, CERN/LHCC 97--32, CMS TDR 3, CERN, 1997.
\bibitem{CMS_zpj}
D.V.~Bandurin, N.B.~Skachkov
``On the application of \zpj events for setting the absolute jet energy scale 
and determining the gluon distribution in a proton at the LHC'', 
hep-ex/0209039.
\bibitem{Owe} 
 J.F.~Owens, Rev.Mod.Phys. {\bf 59} (1987)465.
\bibitem{CMJ}
S.~Abdullin, A.~Khanov, N.~Stepanov, CMS Note CMS TN/94--180 ``CMSJET''.
\bibitem{BKS} 
 D.V.~Bandurin, V.F.~Konoplyanikov, N.B.~Skachkov. 
``Jet energy scale setting with \gpj events at LHC energies. 
JINR Preprints E2-2000-251 -- E2-2000-255, JINR, Dubna.
\bibitem{GPJ_D0}
D.V.~Bandurin, N.B.~Skachkov.
``\gpj process application for setting the absolute  scale of 
jet energy and determining the gluon distribution at the Tevatron Run II.'' D0 Note 3948, 2002.
\bibitem{PYT}
T.~Sjostrand, Comp.Phys.Comm. {\bf 82} (1994)74.
\bibitem{Kot}
A.V.~Kotikov and G.~Parente,
Nucl.Phys.{\bf B549} (1999)242.

\end{thebibliography}
\end{document}